\documentclass[nologo,url,11pt,a4paper]{ETHpaper}
\pdfoutput=1

\usepackage{amsmath, amsthm, amssymb}

\DeclareMathAlphabet{\mathitbf}{OML}{cmm}{b}{it}
\DeclareMathAlphabet{\mathbf}{OML}{cmm}{b}{it}
\usepackage{graphicx}                  
\usepackage{subfig}                    
\usepackage[numbers,sort&compress]{natbib}

\title{Agent-Based Modeling of Intracellular Transport}

\author{Mirko Birbaumer, Frank Schweitzer}

\address{Chair of Systems Design, ETH Zurich, Kreuzplatz 5, 8032 Zurich,
  Switzerland}

\titlealternative{Agent-Based Modeling of Intracellular Transport \\
  \emph{European Physical Journal B} vol. 82, no. 3-4 (2011) pp. 245-255}

\www{\url{http://www.sg.ethz.ch}}

\begin{document}
\maketitle
\renewcommand{\thefootnote}{ \fnsymbol{footnote}} 

\begin{center}
\emph{Dedicated to Werner Ebeling on the occasion of his 75th birthday}  

\end{center}

\begin{abstract}
  We develop an agent-based model of the motion and pattern formation of
  vesicles. These intracellular particles can be found in four different
  modes of (undirected and directed) motion and can fuse with other
  vesicles. While the size of vesicles follows a log-normal distribution
  that changes over time due to fusion processes, their spatial
  distribution gives rise to distinct patterns. Their occurrence depends
  on the concentration of proteins which are synthesized based on the
  transcriptional activities of some genes. Hence, differences in these
  spatio-temporal vesicle patterns allow indirect conclusions about the
  (unknown) impact of these genes.

  By means of agent-based computer simulations we are able to reproduce
  such patterns on real temporal and spatial scales.  Our modeling
  approach is based on Brownian agents with an internal degree of
  freedom, $\theta$, that represents the different modes of
  motion. Conditions inside the cell are modeled by an effective
  potential that differs for agents dependent on their value
  $\theta$. Agent's motion in this effective potential is modeled by an
  overdampted Langevin equation, changes of $\theta$ are modeled as
  stochastic transitions with values obtained from experiments, and
  fusion events are modeled as space-dependent stochastic transitions.
  Our results for the spatio-temporal vesicle patterns can be used for a
  statistical comparison with experiments.  We also derive hypotheses of
  how the silencing of some genes may affect the intracellular transport,
  and point to generalizations of the model.

\end{abstract}

\newcommand{\mean}[1]{\left\langle #1 \right\rangle}
\newcommand{\abs}[1]{\left| #1 \right|}
\section{Introduction}
\label{sec:intro}

Agent-based modeling has proven to be a versatile tool to simulate
processes of structure formation bottom up. By assuming features and
interaction rules of agents on the ``microscopic" level, one is able to
observe the emergent systems properties on the macroscopic level. This is
of particular importance in those areas where the systems dynamics can
hardly be captured top down, i.e. in living systems, including
biological, social or economic systems.

But the advantage of agent-based models in freely defining agent
properties and interactions soon turns out to be a pitfall, because this
way arbitrary patterns can be generated and it is difficult to choose the
right values in a high dimensional parameter space. To minimize these
problems, there are basically two ways: (i) to closely link the agent's
properties to experimentally observed data, and (ii) to apply methods
that allow to aggregate the agent dynamics, to formally derive the
systems dynamics. The latter provides a firm relation between agent's
features and systems feature's and may reveal also the role of certain
(control) parameters.

The concept of Brownian agents \citep{Schweitzer} was developed to
facilitate the second way. The dynamics of agents is described in a
stochastic manner, similar to the Langevin approach of Brownian
motion. This allows to obtain on the macroscopic level a closed-form
partial differential equation for the density, that for the case of
Brownian motion simply describes a diffusion process. The dynamics in
most real systems, however, is much more complicated. Agents are not
simple random walkers, they respond to information in their environment,
follow chemical gradients, and can at the same time also contribute to
generating information, chemical gradients etc. Further, agents do not
behave the same all the time. Instead, they may have different modes of
activity each of which corresponds to a particular behavior.  To cope
with these features, Brownian agents are described by internal degrees of
freedom and their environment is modeled as an adaptive landscape, or
effective potential, which can be modified by the agents while responding
to the information provided. Further, transitions between the agent's
internal degrees are possible, dependent on internal or external
conditions. 

On the formal level, the macroscopic dynamics is then no longer described
by a diffusion equation, but by a quite advanced reaction-diffusion
equation with a variable drift term, which is coupled to another
differential equation describing the dynamics of the adaptive landscape
dependent on the agent's activity.  This allows to tackle the dynamics of
systems comprised of many interacting agents on two levels: (i) the agent
level, where computationally efficient computer simulations can be
performed, (ii) the system's level, where coupled differential equations
may be obtained and investigated analytically.  The application discussed
in this paper is unfortunately complex enough to not provide closed form
equations for an analytical treatment. Nevertheless, the concept of
Brownian agents allows us to formally specify the agent dynamics in terms
of stochastic equations of motion in an adaptive landscape.

The aim of this paper is twofold: (i) to develop an agent-based model of
intracellular transport and pattern formation, which is general enough to
be applied to various such phenomena involving free and directed motion
and fusion processes (Section 3), and (ii) to specify this agent-based
model for the case of vesicle movement and fusion, in close relation to
experimental findings (Section 4).
Importantly, the internal degrees of freedom of agents and transitions
between these are obtained from experiments. This allows us to observe
pattern formation on real time and spatial scales (Section 5), the
outcome of which can, at least in a statistical manner, be compared with
real experiments. Hence, applying the concept of Brownian agents to a
real problem, i.e. the intracellular transport and pattern formation of
vesicles, demonstrates both the versatility of the concept and its
suitability to generate hypotheses about real intracellular processes.

\section{Vesicle Formation and Vesicle Motion}
\label{sec:intromodels}

In this paper, we are interested in the intracellular transport and
pattern formation of \emph{vesicles}. These are quite small intracellular
particles (diameter approx. $0.1\mu$m) (see \citep{roth2006}, \citep{sandy} 
and \citep{mcmahon}). They are formed at the cell
membrane, to contain some extracellular material engulfed by the cell
membrane. This import of material, called endocytic cargo, may include
macromolecules, but also viruses, which are all encapsulated in vesicles
-- a process called \emph{endocytosis} (see \citep{mcmahon}, \citep{marsh2006} 
and \citep{Jason}). In this paper, we do not consider
endocytosis explicitly, but assume that vesicles are formed at the
membrane and then released into the interior of the cell at a constant
rate (later called internalization rate). It is known from experiments
that the size distribution of newly formed vesicles follows approximately
a log-normal distribution (see \citep{dissbirbaumer}). 

\begin{figure}[hbt]
  \centering
  \includegraphics[width=7.5cm]{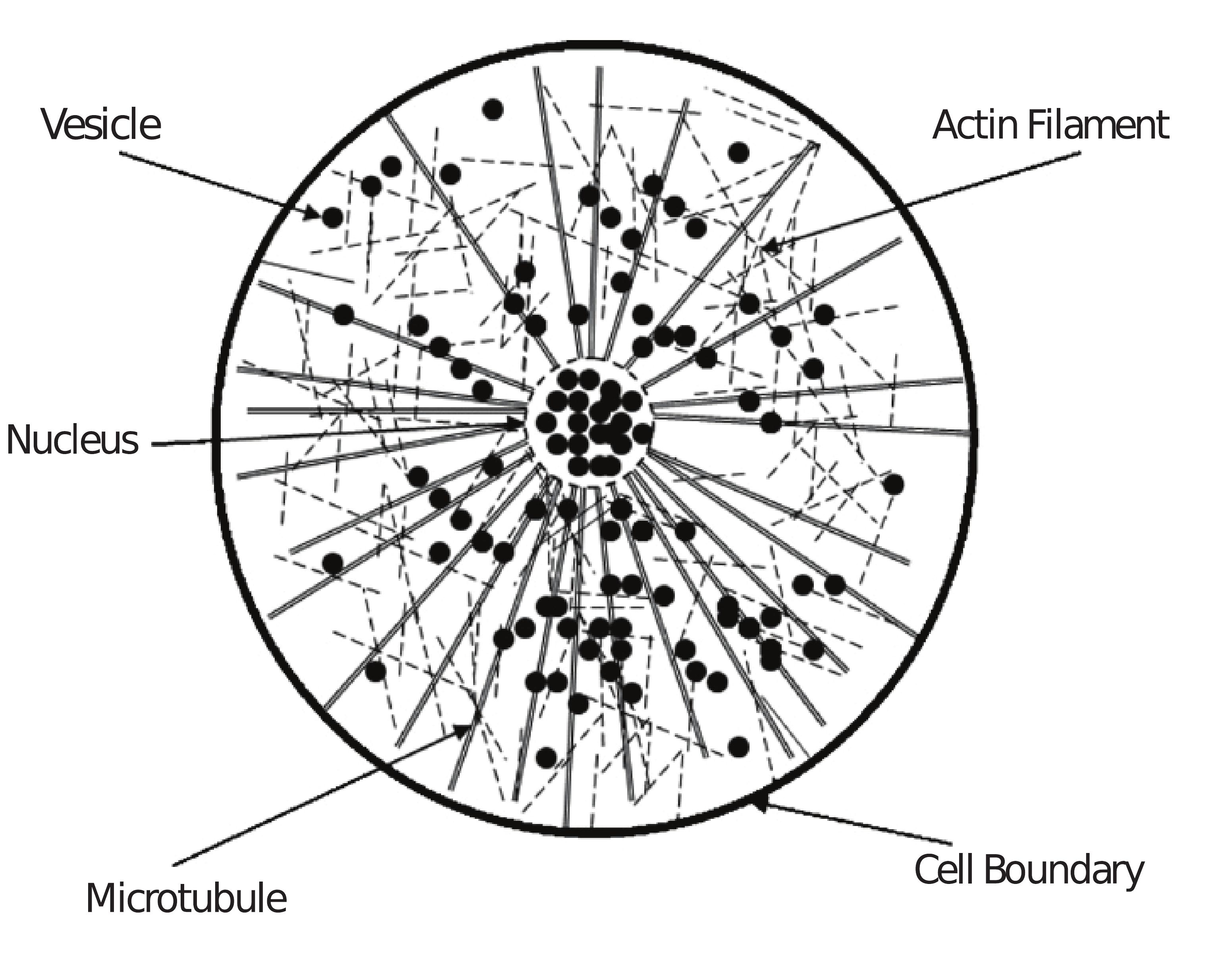}
  \caption{Two-dimensional representation of a cell with cytoskeletal structure. Adapted from \citep{Dinh}.}
  \label{fig:patterns}
\end{figure}

Released vesicles can diffuse inside the cell, but they can also be
reabsorbed by the membrane at a constant rate, later called recycling
rate (\citep{varsat}).  Vesicles need to be transported from the cell membrane to the
endosomal system located in different areas inside the cell -- this
transport process is called \emph{membrane trafficking} (see \citep{mcmahon}).  To facilitate
this process, in addition to the \emph{free diffusion}, vesicles can also
perform a \emph{directed motion} along the \emph{cytoskeleton}, which is
an intracellular structure made of two different kinds of polymer
filaments: actin filaments and microtubule filaments (see Figure
\ref{fig:patterns}). Whereas actin filaments are randomly distributed,
microtubules (MT) are directed toward the microtubule organization center
(MTOC), which is located close to the cell nucleus (see Figure
\ref{modelpatternscontrol}). 
In order to perform a directed motion along
actin or MT filaments, vesicles have to bind to motor proteins
(kinesines, dyneins and myosins) which basically determine the type of
motion. 
The interaction between motor proteins and their related filaments
depends on several additional factors, most notably the presence of
specific proteins such as Rab GTPases, scaffolding proteins and receptor
proteins (see \citep{difiore}). To which extent these proteins are
present depends on the other hand on the genes of the cell, which have to
be transcriptionally active in order to synthesize these proteins -- a
process called \emph{gene expression}.

It is exactly this dependency, which motivates our interest in the motion
of vesicles. If a given gene, for example \emph{cdk8}, is
transcriptionally active, it allows the synthesis of the protein CDK8,
which may affect intracellular processes in various, mostly unknown
ways. In this paper, we precisely ask how such a gene -- through the
synthesis of the specific protein -- affects the motion and pattern
formation of vesicles. The latter process results from the fact that
vesicles can form larger ones by fusing with other vesicles. The fusion
process relies on energy provided by the cell and can only take place on
the MT if vesicles are close enough and below a critical size. The two
simultaneous dynamic processes, namely (free or directed) \emph{motion}
and \emph{fusion} result in a distinct \emph{spatio-temporal}
distribution of vesicles of \emph{different sizes} - which we want to
predict with our model.

The patterns produced by means of our agent-based simulation can then be
compared to experiments which show such vesicle patterns depending on the
transcriptional activity of specific genes (see \citep{dissbirbaumer}). These 
genes can, for example, be knocked out in RNAi or drug screens, which in turn 
perturb the synthesis of proteins (see \citep{hannon2002}). Of course, such 
patterns can be only compared in a statistical sense, a problem discussed in 
the Conclusions. However, if we are able to reproduce empirical patterns with 
our model, we argue that the underlying dynamic processes, motion and fusion, 
are covered sufficiently with our modeling assumptions. This does not only hold for
the assumed interaction between vesicles and MT or other vesicles, it
shall also hold for particular parameter dependencies, most notably the
concentration of specific proteins. Precisely, we want to end up with a
testable prediction of how these concentrations affect vesicle motion and
fusion, which shall be confirmed by subsequent experiments
(see \citep{dissbirbaumer}). Some of the transition rates later used in our model
specifically depend on the experimental setup, e.g. the endocytic
cargo. Here we consider the case of Transferrin, an iron-binding protein
contained in the vesicles to which fluorescently labelled proteins can be
attached, i.e., vesicles can be made visible in the experiment.
The aforementioned internalization rate and recycling rate are thus taken
for Transferrin.

Eventually, we note that our modeling approach (as every modeling
approach) is based on a number of simplifications: (i) we neglect the
motion of vesicles along actin filaments, because it was shown
\citep{Dinh} that such processes do not affect the pattern formation
(recall that fusion only takes place on MT), (ii) we assume that the
cytoskeleton is described by the spatial structure of MT only (i.e. tha
actin filaments are neglected) and that MT are abundant (i.e. there are
always MT to move on) and do not change in time. This allows to neglect
the growth and shrinkage of MT when modeling the motion and fusion of
vesicles.  (iii) We neglect fission, i.e. the fragmentation of larger
vesicles into vesicles of smaller sizes.

\section{A Model of Brownian Agents}\label{sec:Brownian}

\paragraph{Brownian agents}
Our modeling approach is based on the concept of Brownian agents
\citep{Schweitzer} which found many interesting applications in biology
\citep{Schweitzer2, ebeling-fs-03-nova, mach-fs-07, garcia-bacterial-11}
but also in modeling social systems such as online communities
\citep{garcia-fs-10}.  It allows to formalize the agent dynamics using
methods established in statistical physics. A Brownian agent is described
by a set of state variables $u_{i}^{(k)}$, where the index $i=1,...,N$
refers to the individual agent $i$, while $k$ indicates the different
variables.  These could be either \emph{external} variables that can be
observed from the outside, or \emph{internal degrees of freedom} that can
only be indirectly concluded from observable actions.  Noteworthy, the
different (external or internal) state variables can change in the course
of time, either due to influences of the environment, or due to an
internal dynamics.

In the following, each agent represents an intracellular vesicle which,
in accordance with the previous description, is able to change its state
by spatial mobility, changes of activity, and growth processes, as
formalized subsequently.

\paragraph{Spatial mobility}
For the agent's spatial position $r_{i}(t)$, We assume that changes in
the course of time can be described by an overdamped Langevin equation of
a Brownian particle moving in an effective potential \citep{Schweitzer}:
\begin{equation}
  \label{overdampedLang}
  \frac{d \mathbf{r}_{i}}{dt}=\mathbf{v}_{i}[\theta_{i}(t)] 
= \frac{\alpha[\theta_{i}(t)]}{\gamma_{0}}
  \; \frac{\partial h^{e}(\mathbf{r},t,\theta)}{\partial\mathbf{r}}
  +\sqrt{2 D}~ \mathbf{\xi}_{i}(t)
\end{equation}
The overdamped limit implies that the absolute value of the agent's
velocity $v_{i}$ is approximately constant, but the direction may change
due to stochastic influences. Further, $\mathbf{v}_{i}$ implicitly 
depends on the agent's mode of activity, $\theta_{i}(t)$. 

Eqn. (\ref{overdampedLang}) assumes that the agent's motion is influenced
by two different forces, a deterministic one which results from the
gradient of the effective potential, and a stochastic one, which is
assumed to be Gaussian white noise, $\left\langle \mathbf{\xi}_{i}(t)
\right\rangle=0$, $\left\langle \mathbf{\xi}_{i}(t)\mathbf{\xi}_{j}(t') \right\rangle=\delta_{ij}\delta(t-t')$.
The strength of the stochastic force $D=k_{B}T/\gamma_{0}$ determines, in
the spatial case, the diffusion coefficient, with $\gamma_{0}$ being the
friction coefficient.

The deterministic part contains two important ingredients: The effective
potential $h^{e}(\mathbf{r},t,\theta)$ describes the conditions inside
the cell. The response function $\alpha[\theta_{i}(t)]$ depends on the
internal state of the agent, $\theta_{i}(t)$ and determines what
component of the effective potential actually influences the agent.  Both
are specified later after we made clear the notion of the internal state
$\theta$.

\paragraph{Changes of activity}

We assume that the agent's mode of activity $\theta_{i}(t)$ can be changed,
$\omega(\theta'|\theta)$ being the transition rate from state $\theta$
into any other state $\theta'$. In accordance with the literature
\citep{Dinh}, we distinguish five 
different modes of activity of a vesicle as shown in Table
\ref{tab:theta}, each of which is expressed by a different value of the
internal degree of freedom $\theta$.
\begin{table}[h!]
\centering
\begin{tabular}{r|l}
  \hline
  $\theta=0$& Free diffusion in the cytosol \\
  $\theta=1$& Kinesin-driven transport towards MT plus-ends \\
  $\theta=2$& Dynein-driven transport towards MT minus-ends \\
  $\theta=3$& Kinesin/Dynein-driven transport on MT with tendency
  to fusion with other vesicles \\
  $\theta=4$& Bound to the cell membrane \\
  \hline
\end{tabular}
  \caption{Different modes of activities, MT stands for microtubuli}
\label{tab:theta}
\end{table}

While in principle transitions between all states could be assumed, only
a subset of them is biologically relevant. Table \ref{freepar} in Section
\ref{sec:trans} will list those together with their respective value, i.e. the
expected number of transitions per time unit.

\paragraph{Growth and decay}

We assume that fusion, i.e. the coalescence of two vesicles with sizes
$s_{i}$ and $s_{j}$, can be described by a transition rate
$\omega(s_{i+j}|s_{i},s_{j})$ that depends on the internal states
$\theta_{j}$, $\theta_{i}$ of the agents, i.e. their ability to fuse, and
their effective distance $\abs{\mathbf{r}_{i}-\mathbf{r}_{j}}$. As
described above, fusion is only possible for agents with the internal
state $\theta=4$ which is expressed by the Kronecker delta,
$\delta_{\theta_{i},4}$. Further, due to the volume exclusion, agents can
effectively approach each other only up to a distance $d$ (which
represents an average spatial extension of vesicles). The ability to fuse
also depends on the vesicle size because of the energy required for this
process. Because the available fusion energy is limited, it was observed
experimentally that vesicles of a size larger than $s_{\mathrm{max}}$ do
not fuse. This is considered in the transition rates by an additional
exponential cut-off term which becomes effectively zero if one of the
fusing vesicles reaches the maximum size.
This 
leads us to the transition rate for fusion:
\begin{equation}
  \label{eq:fusion}
  \omega(s_{i}+s_{j}|s_{i},s_{j})=\frac{\omega^{s}\delta_{\theta_{i},4}
    \delta_{\theta_{j},4}}{d+\abs{\mathbf{r}_{i}-\mathbf{r}_{j}}}
  \cdot 
\frac{e^{\varepsilon(2s_{max}-s_{i}-s_{j})}}
  {\left[1+e^{\varepsilon(s_{max}-s_{i})}\right]
    \cdot \left[1+e^{\varepsilon(s_{max}-s_{j})}\right]}
\end{equation}
Here $\omega^{s}$ denotes the fusion affinity which depends on the
protein concentration, and $\varepsilon=0.05$ is chosen as a small
number, to increase the cutoff effect.

Little is known regarding the fission process, i.e. the fragmentation of
a vesicle of size $s_{l}$ into two vesicles of sizes $s_{i}$, $s_{j}$
(with $s_{l}=s_{i}+s_{j}$) . Therefore, we assume that the respective
transition rate $\omega(s_{i},s_{j}|s_{i}+s_{j})$ is a constant $w$ equal
for all possible fragmentation processes, which describes spontaneous
fragmentation.  If $w$ is small compared to other transition rates,
fission can be neglected in first approximation.

We note that, because of the fusion process, the total number of
vesicles, is no longer constant. While a conservation of the total mass
$M$ of all vesicles can still be assumed, both the number of vesicles and
their size distribution $P(N_{1},N_{2},...,N_{s}...,t)$ changes over
time. Therefore, we have
\begin{equation}
  \label{eq:conserv}
  M= \sum_{l=1}^{N}s~N_{s}(t)=const.~;~~ \sum_{l=1}^{N}N_{s}(t)\neq const.
\end{equation}

\paragraph{Effective potential}

After the above distinction between the different values for the internal
degree of freedom $\theta$, we can now specify the effective potential
which depends on these states. $h^{e}(\mathbf{r},t,\theta)$ denotes a
scalar potential field that results from the influence of different field
components $h_{\theta}(\mathbf{r},t)$. Each of these components refers to
specific conditions inside the cell. Compared to the time scale involved
in the motion of the vesicles, some of these conditions can be assumed as
constant in time, but varying across space.

With reference to Table \ref{tab:theta}, 
$h_{4}(\mathbf{r})$ describes the influence of the cell membrane on the
motion of agents in state $\theta=4$ as they can bind to the membrane.
$h_{^1}(\mathbf{r})$ and $h_{2}(\mathbf{r})$ determine the agent's motion
along the microtubule filaments.  $h_{0}(\mathbf{r})$ on the other hand
represents the cell topology, i.e. it generates a repelling force close
to the cell membrane and to the nucleus, but is is simply a constant
inside the cell, because free diffusion inside the cell should not be
affected.

The only time-dependent component of the effective field is $h_{3}(\Delta
\mathbf{r},t)$, which affects the fusion processes between vesicles
(fission neglected). In fact, this is a short range attraction potential
which increases with decreasing distance $\Delta
\mathbf{r}=\abs{\mathbf{r}_{i}-\mathbf{r}_{j}}$. Since agents move,
$h_{3}$ changes in time depending on their actual positions
$\mathbf{r}_{i}(t)$ and internal states $\theta_{i}(t)$.

In order to describe how the \emph{effective} potential results from the
different field components, we have to consider the response function
$\alpha[\theta_{i}(t)]$ that determines which of the field components are
actually ``seen'' by the agents conditional on their internal states. In
accordance with the above distinction, we specify:
\begin{align}
  \label{eq:effective}
  \alpha[\theta_{i}(t)]~h^{e}(\mathbf{r},\theta,t)= 
\alpha_{0} h_{0}(\mathbf{r}) 
& + \delta_{\theta_{i},1}~ \alpha_{1} h_{1}(\mathbf{r})
+  \delta_{\theta_{i},2}~ \alpha_{2} h_{2}(\mathbf{r}) 
\nonumber \\
& +  \delta_{\theta_{i},4}~ \alpha_{4} h_{4}(\mathbf{r})
+  \delta_{\theta_{i},3}~ \alpha_{3} h_{3}(\Delta\mathbf{r},t)
\end{align}
The different $\alpha_{k}$ are dimensional constants.  With
eqn. (\ref{eq:effective}) the motion of every agent is specified
according to eqn. (\ref{overdampedLang}). We point out that agents with
$\theta_{i}=0$ behave like simple Brownian particles which are not
affected by any conditions inside the cell, except for the boundary
conditions.

\paragraph{Master equation}

The state of each individual agent is now described by a triple of three
different state variables $\{r_{i}(t),\theta_{i}(t),s_{i}(t)\}$ that can
change in time according to the processes specified above.  The
multi-agent system is thus described by the \emph{grand-canonical}
$N$-particle distribution function
\begin{equation}
P_{N}= P_{N}(\underline{\mathbf{r}},\underline{\theta},\underline{s},t) = 
  P_{N}(\mathbf{r}_{1},\theta_{1},s_{1},...,\mathbf{r}_{N},\theta_{N},s_{N},t),
\end{equation}
which describes the \emph{probability density} of finding $N$ Brownian
agents with the distribution of internal parameters $\underline{\theta}$,
positions $\underline{r}$ and sizes $\underline{s}$ at time
$t$. Note that $N$ is not constant but can change over time due to fusion
(and fission) events. 

The complete dynamics for the ensemble of agents can be formulated in
terms of a \emph{multivariate master equation}: 
\begin{subequations} \label{multi}
  \begin{align}
    \frac{\partial}{\partial t} P_{N}(\underline{\mathbf{r}},
    \underline{\theta}, \underline{s}, t) = & - \sum_{i=1}^{N} \left\{
      \mathbf{\nabla}_{i} \left[
        \frac{\alpha[\theta_{i}(t)]}{\gamma_{0}}~ \mathbf{\nabla}_{i}
        h^{e}(\mathbf{r},t,\theta) P_{N}(\underline{\mathbf{r}},
        \underline{\theta}, \underline{s}, t) \right] - D
      \mathbf{\Delta}_{i} P_{N}(\underline{\mathbf{r}},
      \underline{\theta}, \underline{s}, t) \right\} \label{rr} \\ & +
    \sum^{N}_{i=1} \sum_{\theta_{i}'\neq\theta_{i}} \left[
      \omega(\theta_{i}|\theta_{i}')~ P_{N}(\underline{\mathbf{r}},
      \theta_{i}', \underline{\theta}^{*}, \underline{s},t) -
      \omega(\theta_{i}'|\theta_{i}) P_{N}(\underline{\mathbf{r}},
      \underline{\theta}, \underline{s},t) \right] \label{dd} \\ & +
    \sum^{N}_{i=1} \sum_{i<j} \left[ \omega(s_{i}+s_{j}|s_{i},s_{j})
      P_{N+1}(\underline{\mathbf{r}}^{*}, \theta_{i}=3, \theta_{j}=3,
      \underline{\theta}^{*}, \underline{s}^{*},t) \right. \nonumber \\
    & \left.  - \omega(s_{i},s_{j}|s_{i}+s_{j}) P_{N}(\underline{\mathbf{r}},
      \theta_{i}=3, \theta_{j}=3, \underline{\theta}, \underline{s},t)
    \right] \label{ff}
  \end{align}
\end{subequations}
The first part of the multivariate master eqn., (\ref{rr}), describes
changes of the probability distribution due to movements of agents either
by diffusion or following some gradients of the effective field. The
second part, (\ref{dd}), considers all possible changes of the distribution
of internal states, $\underline{\theta}$, where $\underline{\theta}^{*}$
denotes ``neighboring'' states that differ from $\underline{\theta}$ only
by the element explicitly given. The third part, (\ref{ff}), eventually
describes the fusion process by any two agents. This leads to a ``gain''
if the total number of vesicles is decreased by $N+1 \to N$ and
distribution changes result in $\underline{r}^{*}\to \underline{r}$,
$\underline{s}^{*}\to \underline{s}$, or to a ``loss'' for any other
process with $N\to N-1$.

The multivariate master equation has the advantage of considering all
possible processes on the agent level in a stochastic framework. After
completely specifying the transition rates involved, we are able to solve
this equation by means of stochastic computer simulations (see Section
\ref{sec:sim}). 

We note that from eqn. (\ref{multi}), one can in principle derive a macroscopic
density equation by introducing the agent density of the grand-canonical
ensemble: 
\begin{equation}
  \label{eq:grand}
  n(\mathbf{r},t)= \sum_{N=1}^{\infty} N \int dr_{1}...dr_{N-1} 
P_{N}(r_{1},...,r_{N-1},r,t)
\end{equation}
The resulting equation would have the known structure of a
reaction-diffusion equation for $n(\mathbf{r},t)$. While this may be the
``classical'' way of investigation, we point out that for the agent-based
approach proposed here the stochastic framework is the more appropriate
one because it refers to the individual processes on the agent level. 
We emphasize that the agent-based approach is better suited for
\emph{computer simulations} of spatiotemporal patterns because it
provides a \emph{stable} and \emph{fast} numerical algorithm. This
becomes especially important in the case of large \emph{density
  gradients}, which may considerably decrease the time step allowed to
integrate the related dynamics.

\section{Setup for stochastic simulations}
\label{sec:setup}

\subsection{Velocities and transition rates}
\label{sec:trans}

According to the distinction between the activity modes given in Table
\ref{tab:theta}, the movement of the agents can be of two types:
\emph{free motion} by diffusion processes, described by the diffusion
coefficient $D$, and \emph{bound motion} along the microtubuli. Both
types are described by eqn. (\ref{overdampedLang}). In order to calculate
the different $\mathbf{v}_{i}(\theta)$, we would need to specify explicitly the
related components of the effective potential, $h_{\theta}(\mathbf{r},t)$
that result in the directed motion along the microtubuli. To simplify the
procedure and match it with experimental findings, we may instead
consider the value of $v_{i}(\theta)$ as given by experiments. Then, the role
of the field component is reduced to simply keep the agents on the
microtubuli as long as they are in the respective internal states
$\theta$. I.e. both $\nabla_{i} h_{1}(\mathbf{r})$ and $\nabla_{i}
h_{2}(\mathbf{r})$ just determine the direction of motion along the
microtubulus to which agent $i$ is attached, whereas $h_{0}(\mathbf{r})$
specifies the boundary condition given by the location of the cell 
nucleus and the cell membrane (see also Figure \ref{modelpatternscontrol}).  Table \ref{constpar} 
provides the values for the velocities which are assumed to be constant 
and the same for all agents in the respective internal state.
\begin{table}[htbp]
  \centering
  \begin{tabular}{|c|l|l|}
    \hline
    State &   Parameter value & Description \\
    \hline
    $\theta=0$ &    $D= 10^{-15}~ \rm m^{2}/ s$  & {Diffusion coefficient in cytoplasma}   \\
    \hline 
    $\theta=1$ &
    $v_{1}= 0.33~ \rm \mu m/s$ & {Velocity of vesicle on MT towards plus-ends}   \\
    \hline
    $\theta=2$ &
    $v_{2}=  0.33~ \rm \mu m/s$ & {Velocity of vesicle on MT towards minus-ends}  \\
    \hline
   \end{tabular}
   \caption{\label{constpar}
     Parameters describing the free ($\theta=0$) or bound ($\theta=1,2$)
     motion of agents. MT means `microtubulus'. Note that the diffusion
     coefficient is related to the friction coefficient via
     $\gamma_{0}=k_{B}T/{D}$, which results in $ \gamma_{0}= 4\cdot
     10^{-6} \rm kg/s$. Values according to \citep{mitragotri}.}
\end{table}

As a next step, we need to specify the transition rates
$\omega(\theta'|\theta)$ between different modes of activity. Again,
instead of providing explicit expressions, we may simply take values
obtained from experiments. These values are available to us only for the
unperturbed state of the cell, i.e. for a baseline or reference case in
which conditions inside the cell are not changed on purpose. In the
following, baseline values are indicated by the superscript $b$.  Table
\ref{freepar} lists all biologically relevant transition rates for the
unperturbed scenario, as obtained either from the literature or from own
experiments.
\begin{table}[htpb]
  \centering
  \begin{tabular}{|l|l|}
    \hline
    {Parameter Value} & {Description} \\ 
    \hline
    $\omega^{b}(1|0)$=0.05 $\rm s^{-1}$  &  {Transition from diffusive to MT (plus-end) bound state} \\
    \hline
    $\omega^{b}(0|1)$=0.33 $\rm s^{-1}$  &  {Transition from MT-bound (plus-end) to diffusive state} \\
    \hline
    $\omega^{b}(2|0)$=0.05 $\rm s^{-1}$  &  {Transition from diffusive to MT (minus-end) bound state}    \\
    \hline
    $\omega^{b}(0|2)$=0.33 $\rm s^{-1}$  &  {Transition from MT (minus-end) bound to diffusive state}  \\
    \hline
    $\omega^{b}(3|1)$=0.01  $\rm s^{-1}$ $^{\star}$ &  {Transition from MT-bound to MT-bound state with fusion} \\
    \hline
    $\omega^{b}(1|3)$=0.02  $\rm s^{-1}$  $^{\star}$&  {Transition from MT-bound state with fusion to MT-binding}  \\
    \hline
    $\omega^{b}(3|2)$=0.01  $\rm s^{-1}$   $^{\star}$ &  {Transition from MT-bound to MT-bound state with fusion} \\
    \hline
    $\omega^{b}(2|3)$=0.02  $\rm s^{-1}$  $^{\star}$ &  {Transition from MT-bound state with fusion to MT-binding}  \\
    \hline
    $\omega^{b}(0|4)$=0.0025-0.0033 s$^{-1}$  &  {Internalization rate of vesicles} \\
    \hline
    $\omega^{b}(4|0)$=0.00083-0.0012 s$^{-1}$  &  {Recycling rate of vesicles} \\
    \hline
  \end{tabular}
  \caption{\label{freepar}
    Biologically relevant transition rates $\omega^{b}(\theta'|\theta)$ for the
    baseline case (unperturbed concentrations of proteins). Values
    according to \citep{mitragotri}, values with $^{\star}$ are from own
    experiments, see \citep{dissbirbaumer}. }
\end{table}

\subsection{Reduced transition rates}
\label{sec:reduced}

The given set of transition rates leaves us with a large degree of
freedom which however turns out to be a pitfall: if we wanted to compare
the simulations with patterns observed in the experiment, we would need
to raster the full parameter space to find appropriate combinations of
transition rates which lead to a realistic outcome. There are basically
two ways to reduce this parameter space: (i) we use known values of the
transition rates as e.g. reported in the literature, (ii) we introduce
reduced transition rates, assuming that not all processes are really
independent.
We follow a combination of the two, defining the reduced transition rates
as given in Table \ref{tab:omega-short}
\begin{table}[h!]
  \centering
\renewcommand{\arraystretch}{2}
  \begin{tabular}{|l|l|}
    \hline 
    $\Omega_{1}=\frac{\displaystyle \omega(1|0)}{\displaystyle \omega(0|1)}$
    &
    \rm Affinity for microbule plus-ends
    \\ \hline
    $\Omega_{2}=\frac{\displaystyle \omega(2|0)}{\displaystyle
      \omega(0|2)}$&
    \rm Affinity for microbule minus-ends
    \\ \hline
    $\Omega_{3}=\frac{\displaystyle \omega(3|1)}{\displaystyle
      \omega(1|3)}$&
    \rm Fusion tendency on microbule plus-ends
    \\ \hline
    $\Omega_{4}=\frac{\displaystyle \omega(3|2)}{\displaystyle
      \omega(2|3)}$&
    Fusion tendency on microbule minus-ends
    \\ \hline
    $\Omega_{5}=\frac{\displaystyle \omega(0|4)}{\displaystyle
      \omega(4|0)}$&
    Internalization versus recycling rate
    \\ \hline
\end{tabular}
\caption{Reduced transition rates $\Omega_{i}$, to combine two
  transition rates $\omega$ which refer to the same intracellular (transport)
  mechanism but describe inverse processes.
}
\label{tab:omega-short}
\end{table}

In the following we assume that $\Omega_{3}=\Omega_{4}\equiv\Omega_{F}$
which denotes the fusion tendency that is supposed to be independent of
the kind of motor protein involved (this was kinesin for $\Omega_{3}$ and
dynein for $\Omega_{4}$). With this, we finally have reduced the
transition rates to 4 parameters given by
$\Omega_{1},\Omega_{2},\Omega_{F}$ and $\Omega_{5}$.

These reduced parameters of course do not fully determine the individual
transition rates which are needed to recover the correct
dynamics. Therefore, for one of the transition rates we use the baseline
value from the literature, as given by Table \ref{freepar}. So, we define
e.g. $\Omega_{1}= \omega(1|0)/\omega^{b}(0|1)$ or $\Omega_{1}=
\omega^{b}(1|0)/\omega(0|1)$.

Eventually, we want to emphasize the distinction between the reduced
transtion rates $\Omega_{i}^{u}$ for the unperturbed case (control cell) and
$\Omega^{k}_{i}$ for the \emph{perturbed} case, where the protein $k$ was
manipulated. Similar to the discussion in Section \ref{sec:trans}, they
refer to the same transition rates $\omega$, but with different
concentrations $c_{k}$.

\subsection{Boundary and initial conditions}
\label{sec:initial}

\begin{figure}[htbp]
  \centering
  \includegraphics[width=5.8cm]{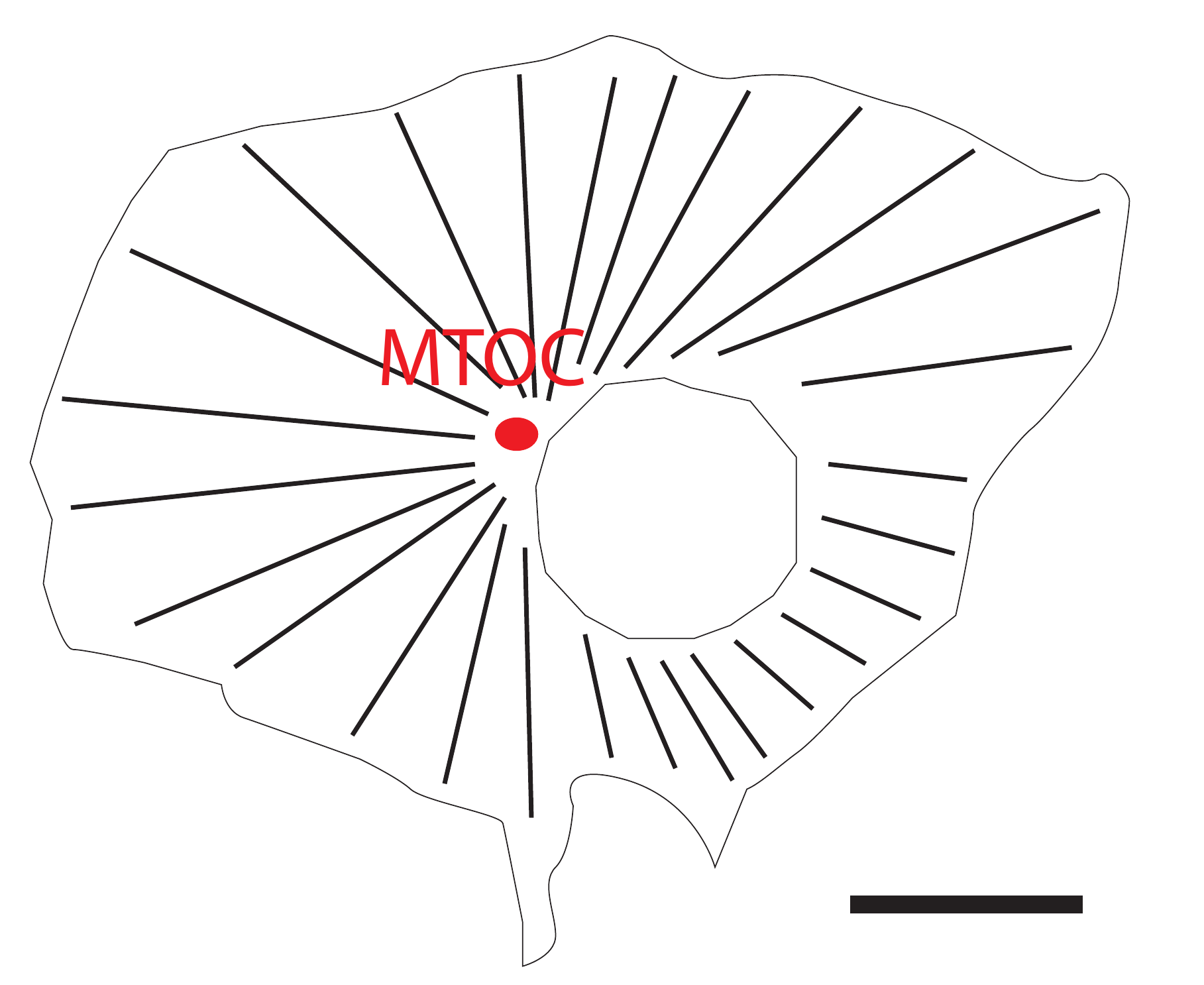}\hfill
  \includegraphics[width=5.8cm]{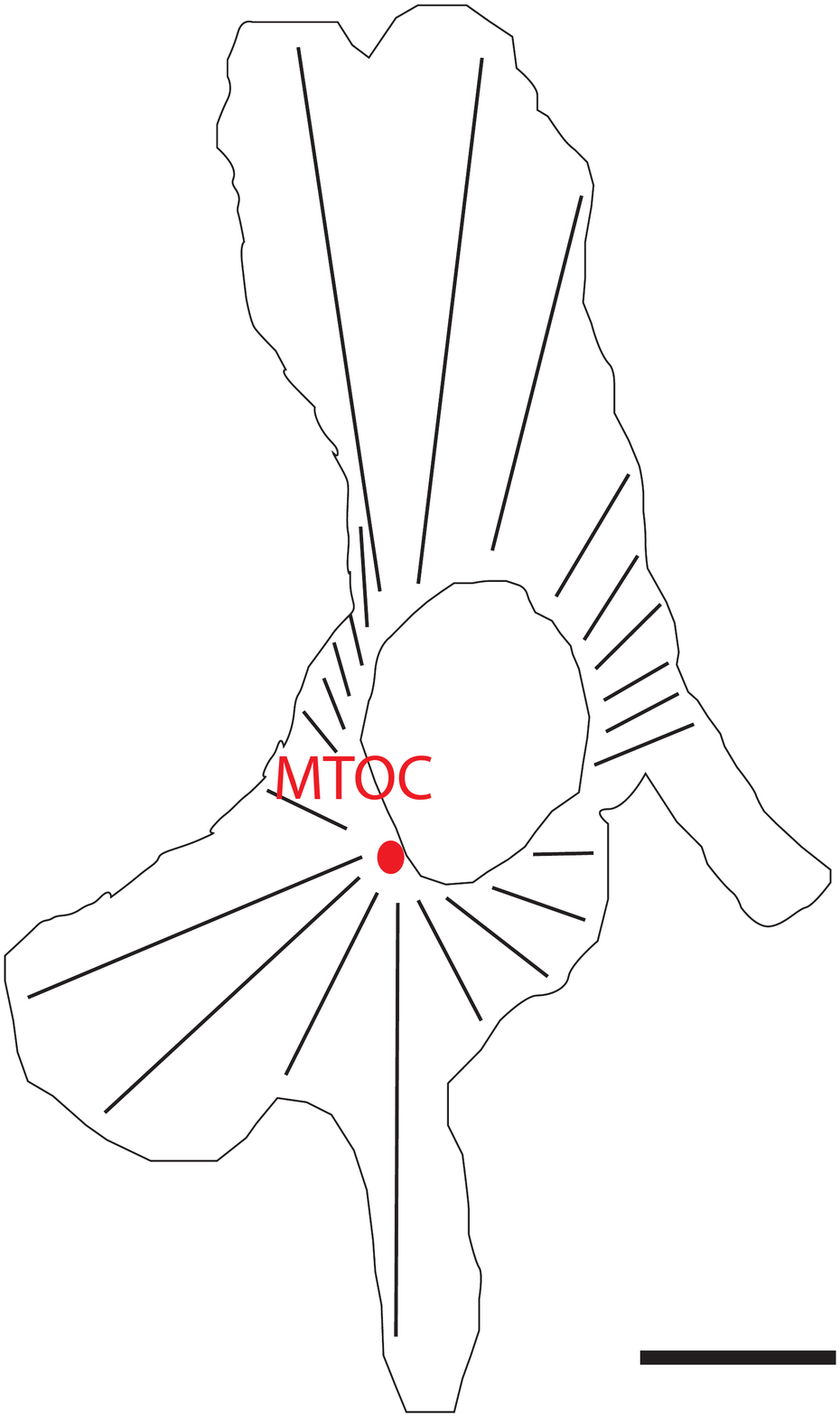}
  \caption{(left) \emph{(a)} Circular geometry of a non-treated (control)
    cell; (right) \emph{(b)} Elongated geometry of a perturbed cell
    (treated with siRNA, which leads to the depletion of CDK8).  The
    scale bar corresponds to 5$\mu m$. The microtubule filaments are
    schematically sketched, pointing toward the MTOC (microtubule
    organization center). Vesicle in an internal state $\theta\in\{1,2,3\}$
    move along microtubules in both directions.}
  \label{modelpatternscontrol}
\end{figure}

In order to complete the setup for the stochastic computer simulations,
we still need to specify the boundary conditions which refer to the cell
geometry. Figure \ref{modelpatternscontrol} (a,b) shows the two different
cell geometries chosen, a rather circular cell and an elongated one. The
outer boundary of the cell membrane and the inner boundary of the nucleus
are both assumed to be impermeable walls, described by a hard sphere
potential $h_{0}(\mathbf{r})$.

The interior of the cell contains the cytoskeleton, i.e. both actin
filaments and microtubuli (MT) which provide boundary conditions for the
motion of vesicles (see also Figure \ref{fig:patterns}). As already
discussed, we do not consider motion along actin filaments because 
the related processes do not contribute to vesicle pattern formation. MT, 
on the other hand, are abundant and always point to the microtubule organization
center (MTOC) which is assumed to be in the perinuclear area on the side
with the largest part of the cytosol (see Figure
\ref{modelpatternscontrol} a,b).

Regarding the initial conditions, we need to specify the \emph{number},
the \emph{position}, the \emph{internal state} and the \emph{size} of the
agents at $t=0$. For our model, we assume that initially all vesicles are
bound to the cell membrane, i.e. at $t=0$ agents start with
$\theta_{i}(0)=4$ at a position $\mathbf{r}_{i}(0)$ randomly chosen from
the cell boundary.  In order to determine the initial number of agents,
we start from the experimental observation that an average cell (of the
type considered) contains about 200 internalized vesicles 
 at steady state. This number excludes vesicles still bound to the 
cell membrane. We further know from experiments that internalization rates, i.e., 
transition rates from the initial bound into the free moving state (and vice versa), 
i.e. $\omega(0|4)=3.3\cdot 10^{-3}\rm s^{-1}$ and 
$\omega(4|0)= 8.3\cdot 10^{-4}\rm s^{-1}$ (see Table \ref{freepar}). 
If $N(0)=N_{4}(0)$ denotes the (unknown) number of agents
at $t=0$ (all assumed to be in the bound state) and $N(0)-N_{4}^{st}=200$
the (known) number of agents no longer in the bound state at steady state
($st$ : steady state), then we can postulate for the dynamics for $N_{4}(t)$ 
the following rate equation:
\begin{equation}
\label{internalization}
 \frac{d N_{4}(t)}{dt}=- \omega(0|4)~ N_{4}(t) + \omega(4|0)~
 \left[N_{4}(0)-N_{4}(t)\right]
\end{equation}
After a time $t\sim 1/[\omega(0|4)+ \omega(4|0)]$, this dynamics reaches
a steady-state solution
\begin{equation}
  \label{eq:ss}
  N_{4}^{st}=\frac{\omega(4|0)}{\omega(0|4)+\omega(4|0)} N_{4}(0)
\end{equation}
from which we can calculate $N_{4}(0)$ assuming that
$N(0)-N_{4}^{st}=200$. This approximation neglects the fact that the
total number of vesicles have decreased at time $t$ because of the fusion
between vesicles. Thus, we may slightly increase the initial number of
agents and have eventually chosen $N_{4}(0)=350$.

It remains to determine the initial distribution of vesicle sizes. We
want to start from a most realistic one because (a) later we want to
compare the time scale of structure formation with the experimental
observation, and (b) because our modeling setup has neglected the
fragmentation rate of vesicles (which would be needed if an arbitrary
initial distribution needs to relax into a realistic one). Again, we rely
on experimental observations \citep{dissbirbaumer} that have found a log-normal
distribution of vesicle sizes: 
\begin{equation}
\label{ln}
  P(s;\mu,\sigma)=\frac{1}{s\sigma\sqrt{2\pi}}\exp{
\left\{-\frac{(\ln s-\mu)^2}{2\sigma^2}\right\}},
\end{equation}
where $\mu$ and $\sigma$ are the mean and standard deviation of the
variable's natural logarithm.

\subsection{Stochastic Simulation Technique}
\label{sec:sim}

We now have determined all ingredients for stochastic computer
simulations which include the following dynamic processes (specified on
the agent level):

\begin{description}
\item[Initialization] At $t=t_{0}$, 350 agents with $\theta_{i}(0)=4$ are
  randomly placed at the cell boundary. Their initial size $s_{i}(0)$ is
  drawn from the log-normal distribution, (\ref{ln}).

\item[Movement] Agents can change from the bound state into the free
  moving state at a rate $\omega(0|4)$ and from the free moving state
  into directed motion at rates $\omega(1|0)$, $\omega(2|0)$. They all
  move according to the equation of motion, (\ref{overdampedLang}).

  In our modeling approach, only agents with the internal states
  $\theta\in\{1,2,3\}$ move along the MT. We assume that, whenever an
  agent switches from $\theta=0$ into either $\theta=1$ or $\theta=2$,
  i.e. from a free motion into a bound motion, a MT is ``always at
  hand''. If the agent switches from $\theta\in\{1,2\}$ into $\theta=3$
  where it is ready to fuse, it continues to move into the same direction
  as before, until it collides with another agent.

\item[Fusion] Precisely, fusion occurs only on MT. Agents need to be in
  state $\theta=3$ and in a sufficiently close distance. After fusion,
  the smaller agent ``disappears'', whereas the larger one has increased
  its size, but keeps the internal state $\theta=3$ until one of the
  transitions $\omega(2|3)$ or $\omega(1|3)$ happen.

\item[Inversion] Agents bound to the MT can become freely moving at the
  rates $\omega(0|2)$, $\omega(0|1)$, whereas free moving agents can be
  bound again to the cell membrane at the rate $\omega(4|0)$. 

\item[Concurrency] We assume that agents can move and transit into
  different internal states at the same time. This allows us to decouple
  the motion of agents from the various reactions (change of internal
  states and fusion).
\end{description}

The state of the multi-agent system is at any time completely described
by the $N$-particle distribution function $P_{N}(\underline{\mathbf{r}},
\underline{\theta}, \underline{s},t)$. However, because of the dynamical
processes, the system state always changes and its average ``life time''
$\mathcal{T}_{m}$ is just the inverse of the sum over all possible transition
rates that can change the given state (including changes of position,
internal states, and sizes).

Because we have to deal with movement and reactions at the same time, we
have chosen a sufficiently small fixed time interval $\Delta t=0.02 s$ to
solve the equations of motion of each individual agent. To answer the
question if in the respective time interval also a change of the internal
states or a fusion process occurs, we proceed as follows:

Each of the possible reactions has a different probability to occur,
which is determined by the respective transition rate and the available
time, $\Delta t$. In order to pick one of the possible reactions, we draw
a uniformly distributed random number $U\in\textrm{RND}[0,1]$ and choose
the process $z$ that satisfies the condition
\begin{equation}
  \sum_{j=0}^{z}\omega_{z}(\cdot)\Delta t < U < 
  \sum_{j=0}^{z+1}\omega_{z}(\cdot)\Delta t
\end{equation}
Because $\Delta t$ was chosen such that $\sum_{n=1}^{N}
\omega_{n}(\cdot)\Delta t \leq 1$, none of the possible processes is
excluded from being picked. On the other hand, it may occur that none of
the processes is being chosen if the sum is much smaller than 1 and $U$
close to 1. Then no reaction occurs during the respective time interval,
but movements take place.

\section{Results}
\label{results}

\subsection{Spatio-temporal vesicle patterns}
\label{sec:pattern}

\begin{figure}[htbp]
  \mbox{} {\bf \hspace{0.8cm} 1 min \hfill 5 min \hfill 10 min \hfill 15 min
\hspace{0.6cm} \mbox{} } \\[1ex]
  \centering
  \includegraphics[width=0.2\textwidth]{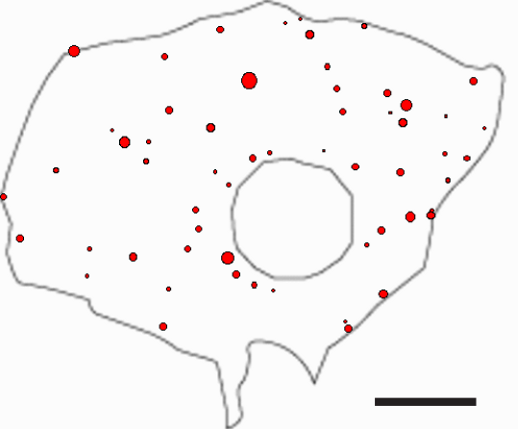}\hfill
  \includegraphics[width=0.2\textwidth]{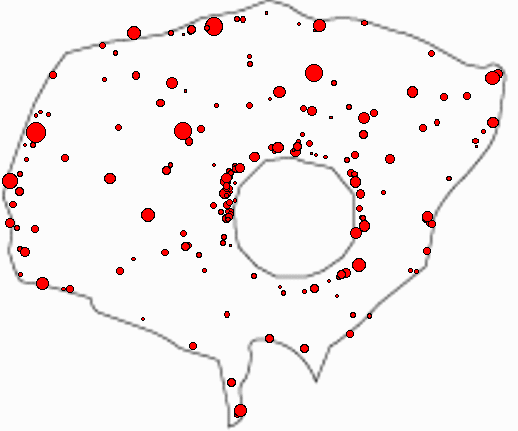}\hfill
  \includegraphics[width=0.2\textwidth]{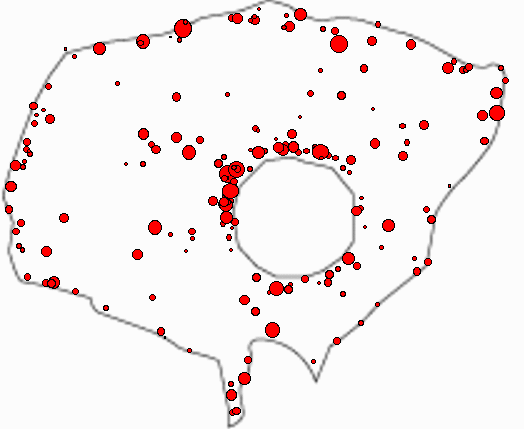}\hfill
  \includegraphics[width=0.2\textwidth]{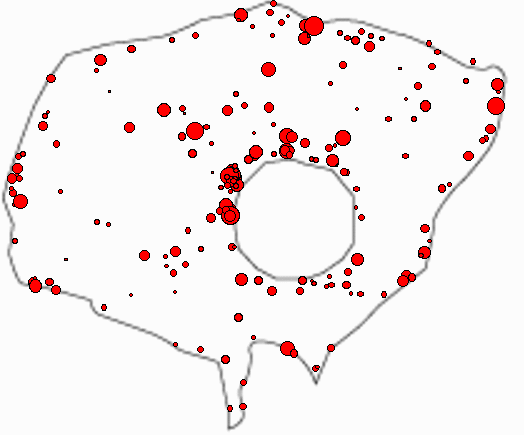}
  \centering
  \includegraphics[width=0.2\textwidth]{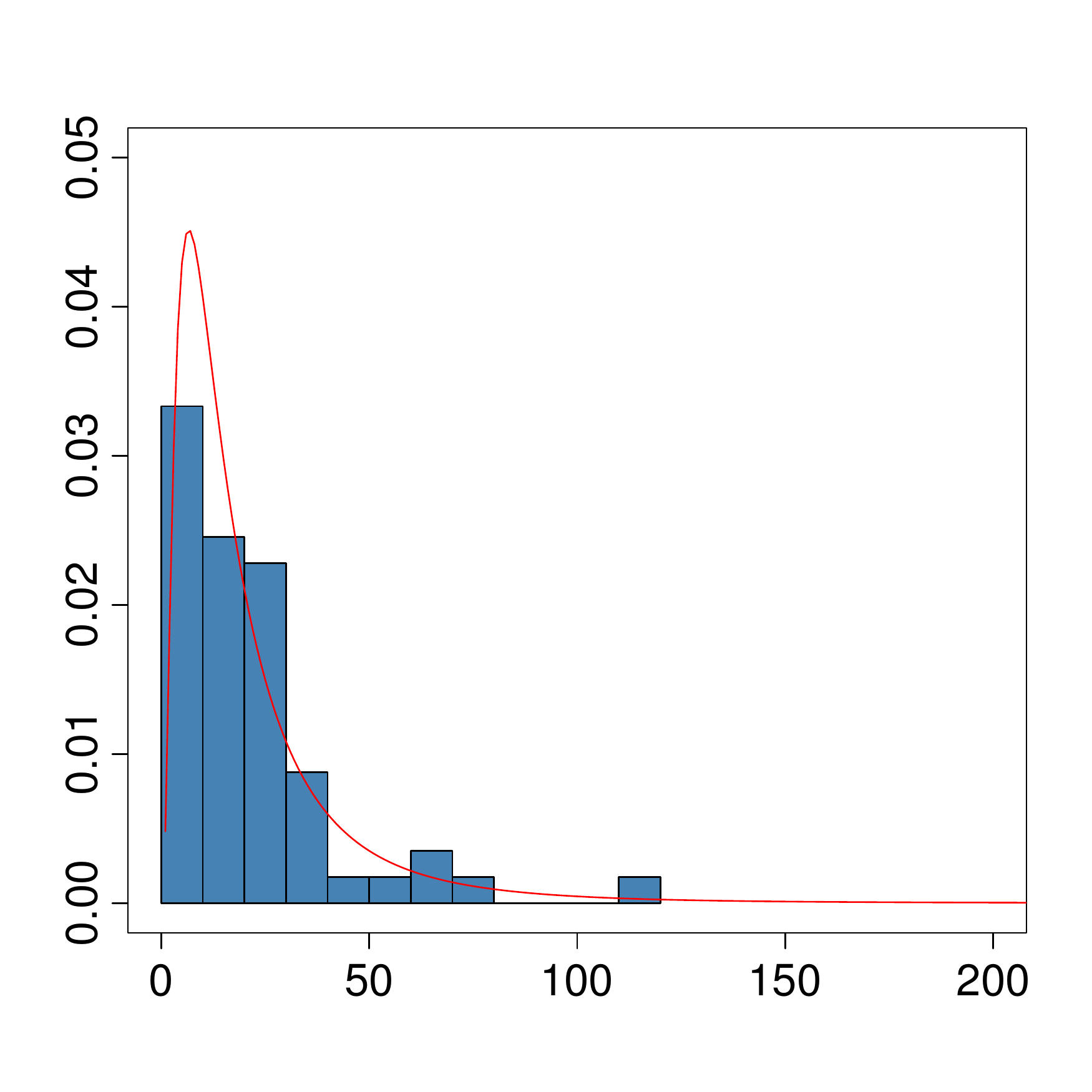}\hfill
  \includegraphics[width=0.2\textwidth]{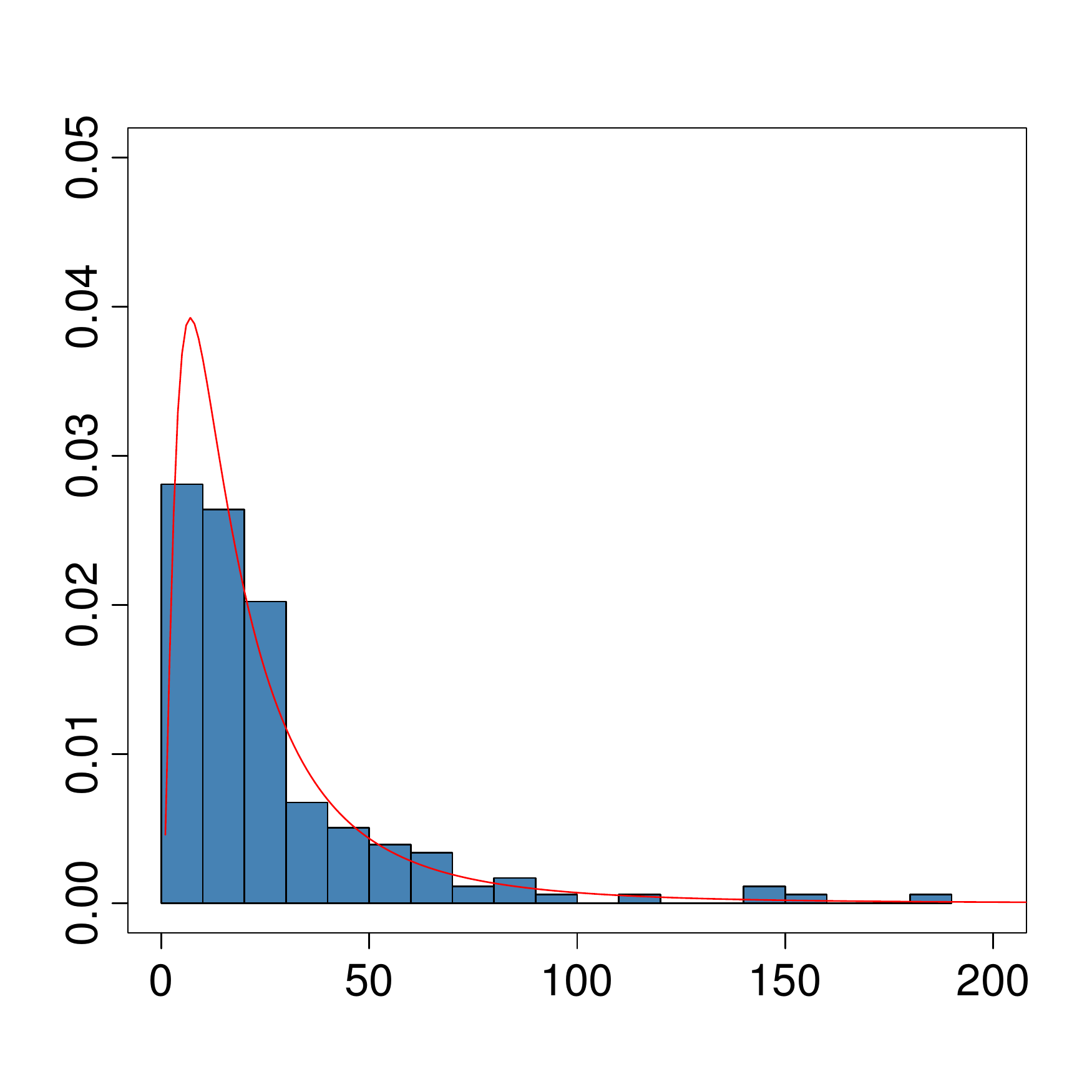}\hfill
  \includegraphics[width=0.2\textwidth]{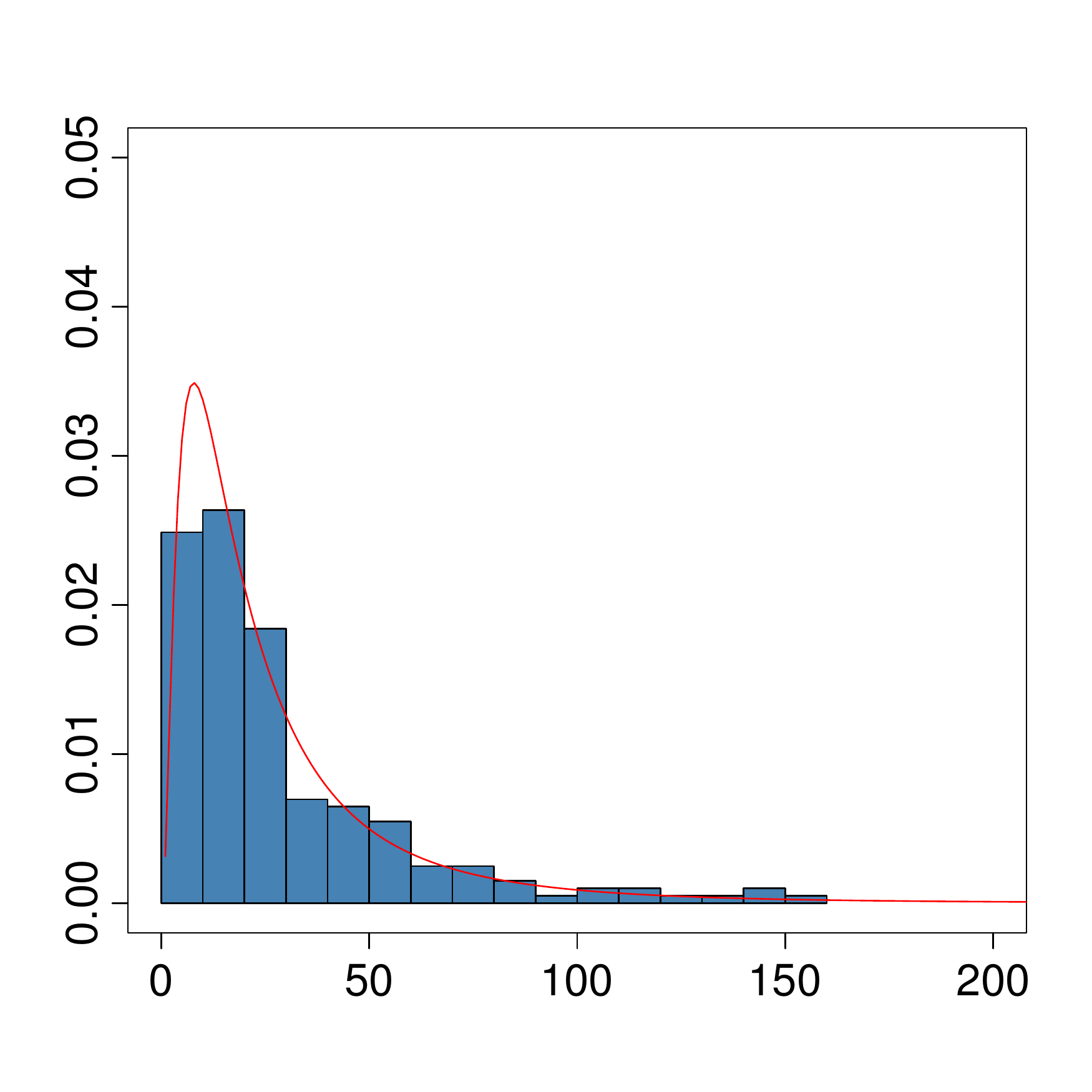}\hfill
  \includegraphics[width=0.2\textwidth]{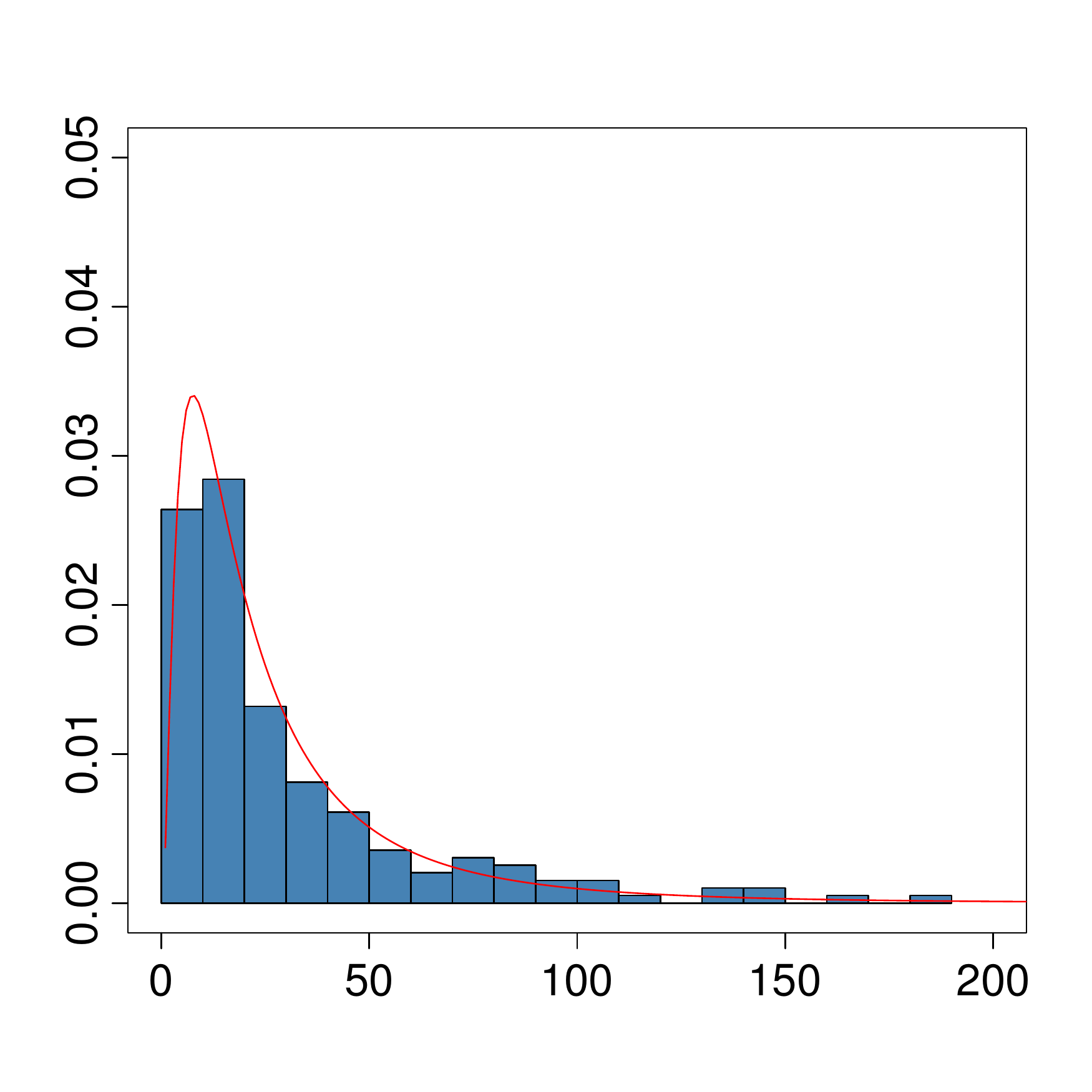}
\medskip \\
  \centering
  \includegraphics[width=0.2\textwidth]{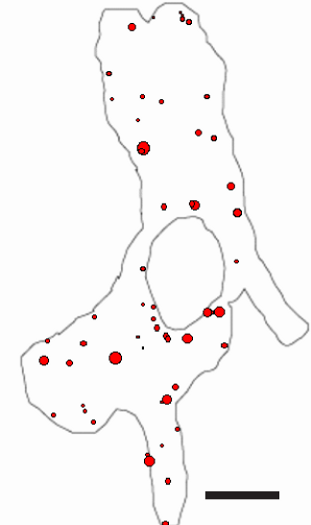}\hfill
  \includegraphics[width=0.2\textwidth]{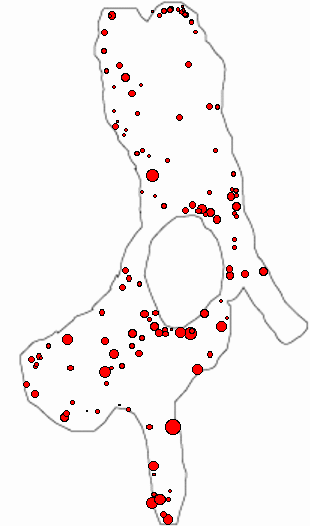}\hfill
  \includegraphics[width=0.2\textwidth]{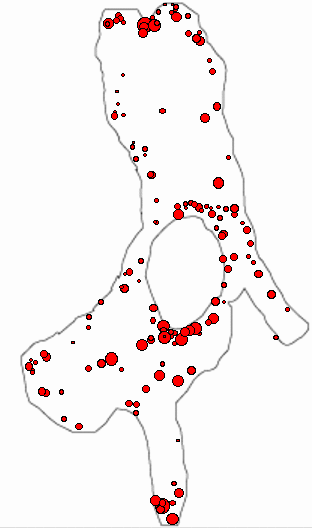}\hfill
  \includegraphics[width=0.2\textwidth]{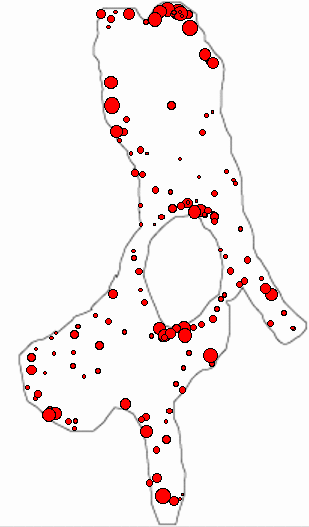}
  \centering
  \includegraphics[width=0.2\textwidth]{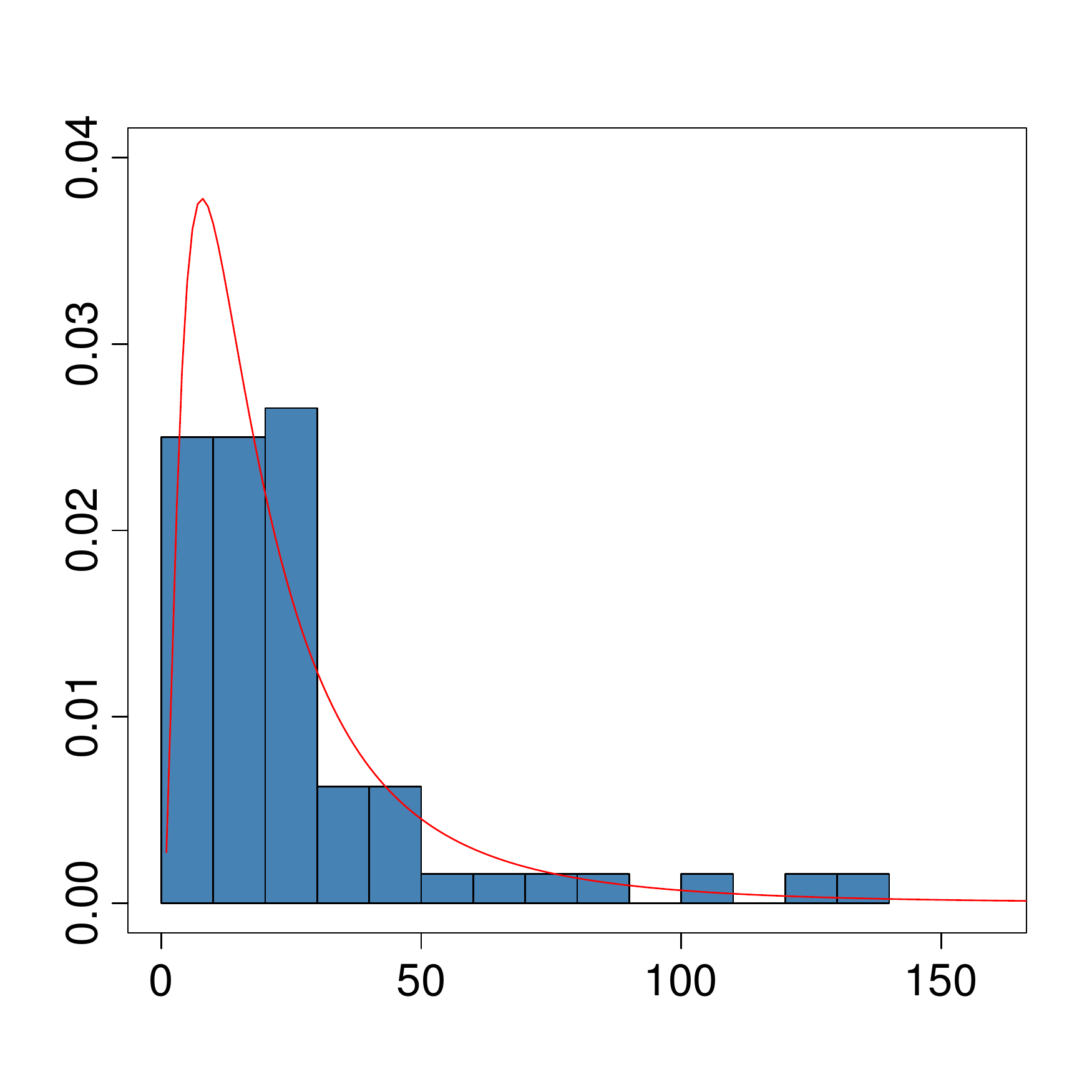}\hfill
  \includegraphics[width=0.2\textwidth]{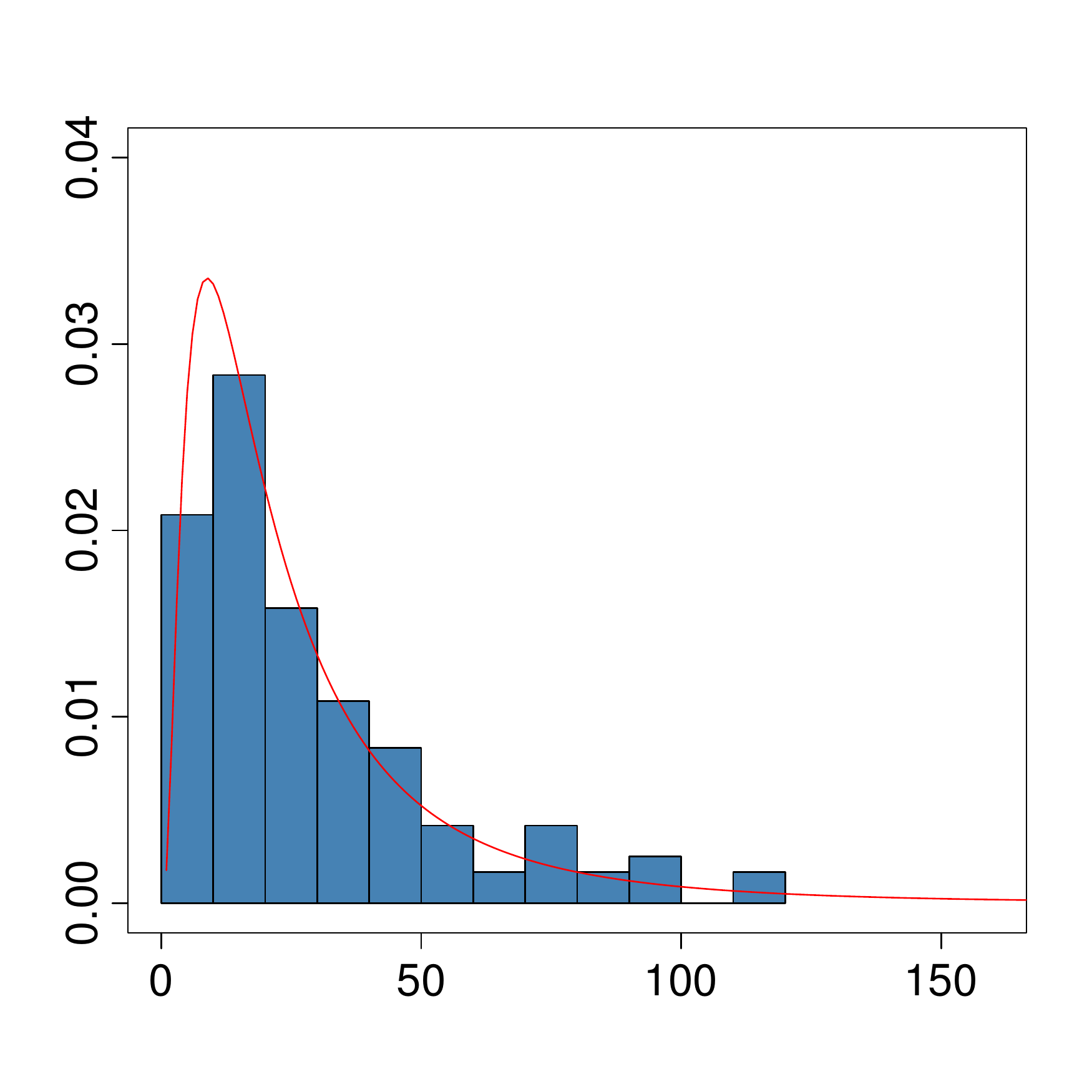}\hfill
  \includegraphics[width=0.2\textwidth]{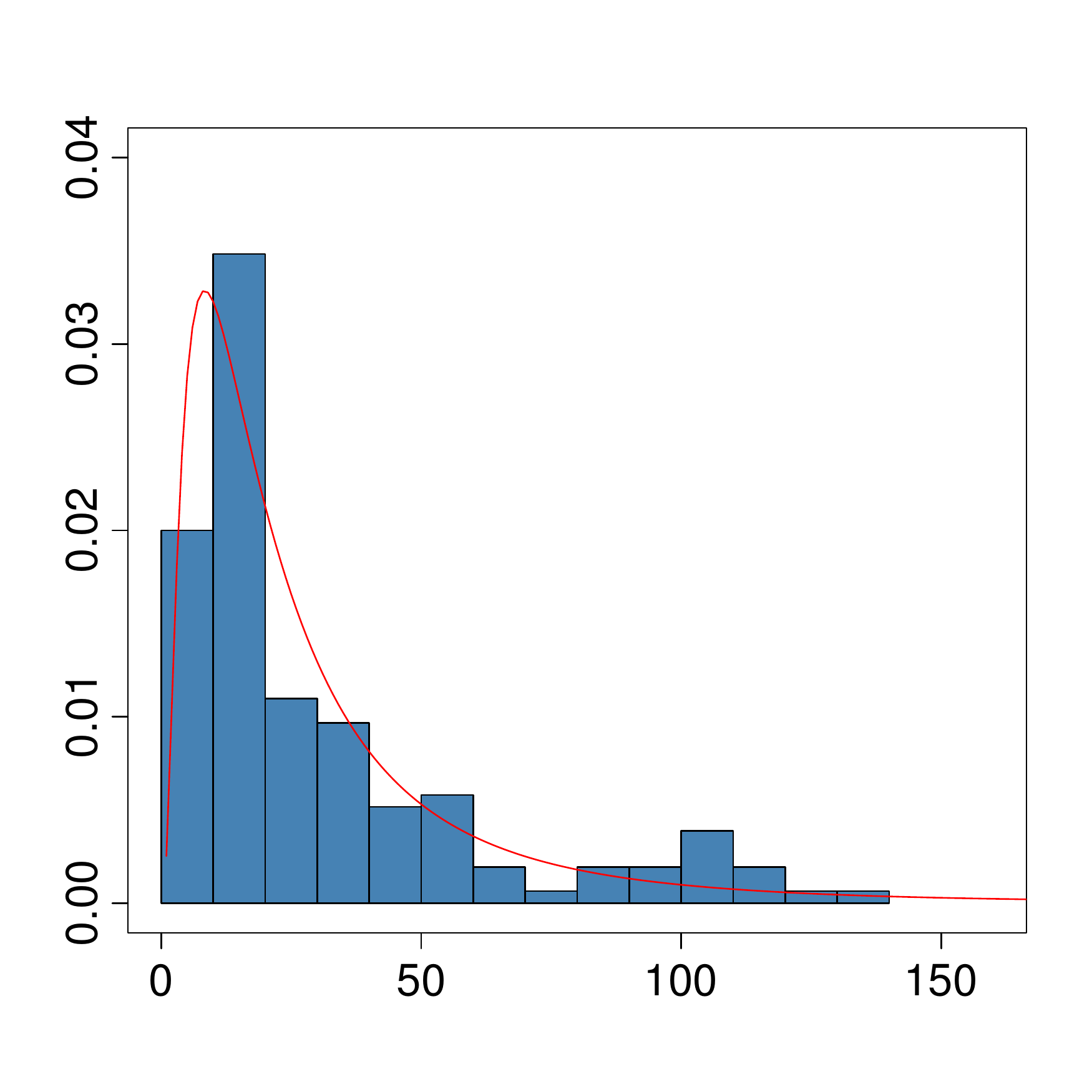}\hfill
  \includegraphics[width=0.2\textwidth]{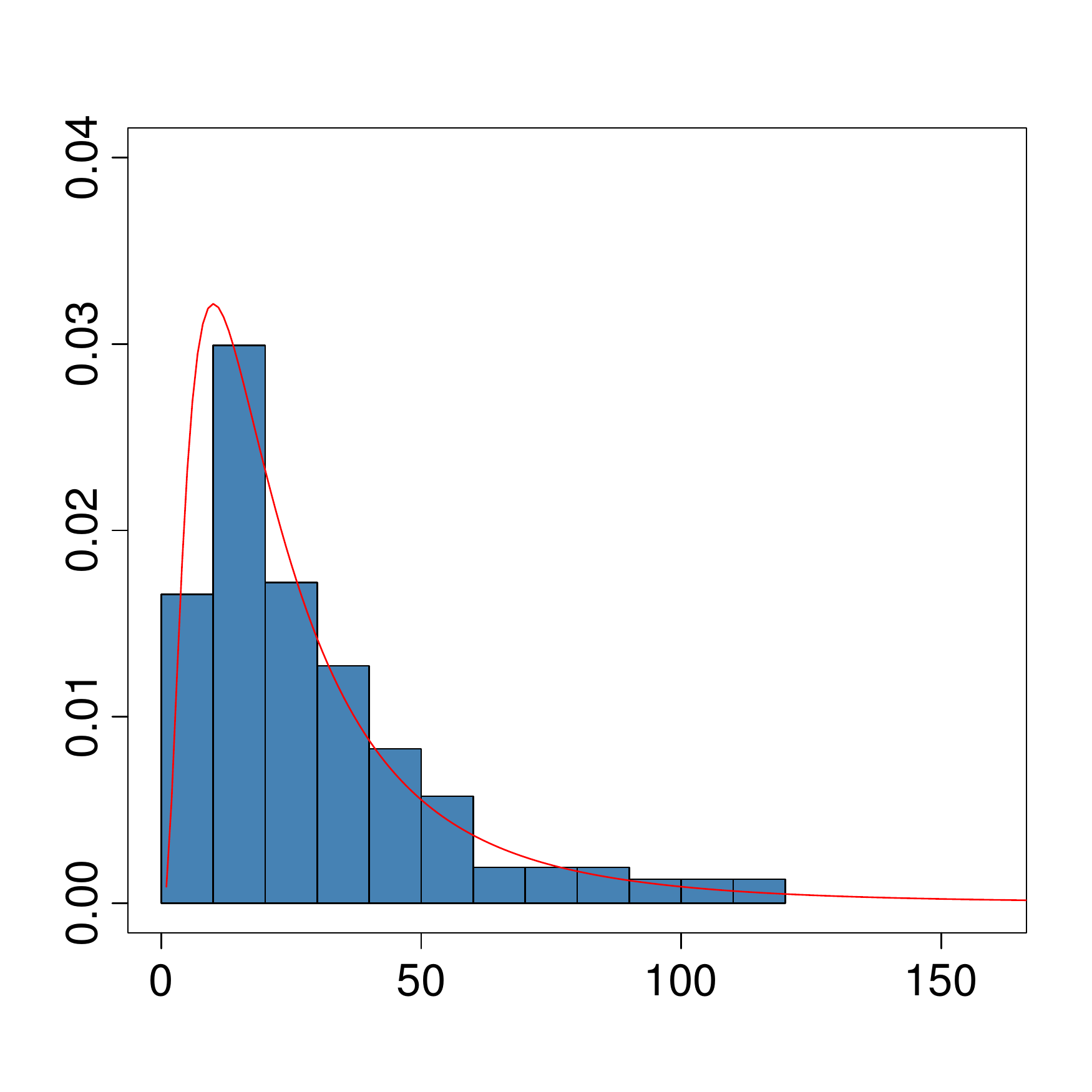}
  \caption{
Computer simulations of vesicle patterns on real time scale (in
    \textrm{min}) and spatial scale (scale bar: 5$\mu$m ). 
(top) unperturbed cell:
$\{\Omega_{1},\Omega_{2},\Omega_{F},\Omega_{5}\}$=$\{0.925, 1.1, 1.0,
0.85\}$, (bottom) per\-turbed cell: $\{\Omega_{1},\Omega_{2},\Omega_{F},\Omega_{5}\}$=$\{1.3, 1.2, 1.2,
0.55\}$. The histograms show the evolution of the vesicle size
distribution together with the fitted log-normal distribution (red
line). Values for $\mu$ and $\sigma$ are given in Table \ref{tab:log}. 
}
\label{fig:patternscdk8}
\end{figure}

Figure \ref{fig:patternscdk8} presents computer simulations of the
vesicle patters for both the unperturbed (control) cell and the perturbed
cell (cf also Figure \ref{modelpatternscontrol}). We emphasise that these
simulations refer to real spatial and time scales, so they should be
comparable, at least in a statistical sense, to patterns observed from
experiments. These experiments are reported in \citep{dissbirbaumer} and
have motivated the choice of the reduced transition rates, $\Omega_{i}$,
which are treated in this paper as free parameters.

Comparing the pattern formation in the perturbed cell with the one in the
control cell, we note a number of differences: we observe a localization
of large vesicles on the one hand at the tips of the elongated
branched-out perturbed cell, on the other hand large vesicles are as well
localized in the perinuclear area of this cell. In the unperturbed cell,
the majority of large vesicles are located around the nucleus and
vesicles are spread over the entire cell surface getting more sparse
towards the cell periphery. 

Comparing the vesicle size distribution of both the perturbed and the
unperturbed cell, we find that they preserve the form of the log-normal
distribution, but the mean value $\mu$, in the course of time, shifts to
significant larger values in the perturbed cell (see also Table
\ref{tab:log}). The perturbed cell displays a geometry that may have
facilitated fusion of vesicles in its branches within which they
accumulate. The transport of vesicles from these branches to the nucleus
of the perturbed cell seems to be suppressed.
\begin{table}[htbp]
  \centering
  \begin{tabular}[c]{|c|c|c|c|c|}
    \hline
    $t$& 1 \textrm{min} & 5 \textrm{min} & 10 \textrm{min} & 15 \textrm{min}
    \\ \hline \hline
    $\mu^{u}$ & 2.69 & 2.82 & 2.94 & 2.96 \\
    $\sigma^{u}$ & 0.89 & 0.94 & 0.94 & 0.96 \\ \hline \hline
    $\mu$ & 2.87 & 2.99 & 3.00 & 3.04 \\
    $\sigma$ & 0.90 & 0.90 & 0.94 & 0.86 \\ \hline
  \end{tabular}
  \caption{Mean $\mu(t)$ and standard deviation $\sigma(t)$ of the
    log-normal distribution at different times $t$ (\textrm{min}) for the
    unperturbed (superscript $u$) and the perturbed cell. }
  \label{tab:log}
\end{table}

\subsection{Estimating concentration dependence}
\label{sec:concentr}

As pointed out above, we found very realistic vesicle patterns both for
the perturbed and the control cell for the given set of reduced
transition rates $\Omega_{i}$ (see Figure \ref{fig:patternscdk8}). These
transition rates are treated as free parameters in our model -- but
provided they are correct (which can only be confirmed by comparing the
patterns with experiments, statistically) they allow an indirect
estimation of the concentration dependence of the transition rates
$\omega(\theta'|\theta)$.

As already stated, the transition rates are given to us only for the
unperturbed state of the cell.  Table \ref{freepar} presents the values
for the baseline case.  One should note, however, that the baseline value
not necessarily describes the experimental situation because it was
obtained under conditions which are hardly reproducible completely.  If
we for example change the concentration $c_{k}$ of some protein ${k}$
inside the cell which is involved in processes of fusion or directed
motion, this will certainly change the value of the respective transition
rates, i.e. $\omega=\omega(c_{k},c_{l},...)$. Hence, it is not only
sufficient to know the values of the baseline case, we should also know
how these values of the transition rates change with the concentrations
$c_{k}$. If we denote the transition rates in the unperturbed case by
$\omega^{u}$ (omitting the $\theta$ dependence at the moment) and the
respective protein concentrations by $c^{u}$ we may assume in first-order
approximation the following expansion:
\begin{equation}
  \label{eq:exp}
  \omega(c_{k},c_{l\neq k}=c^{u}_{l\neq k})= \omega^{u}(c_{k}^{u},c^{u}_{l\neq k})
  + \left. \frac{\partial\omega}{\partial
      c_{k}}\right|_{c_{k}^{u}} (c_{k}-c_{k}^{u}).
\end{equation}
To further specify the functional dependency $\omega(c_{k})$ we make the
following ansatz:
\begin{equation}
  \label{ansatzexp}
  \omega(c_{k},c_{l\neq k}=c^{u}_{l\neq k})= \omega^{u} \cdot
  \exp\left(\kappa_{k} \frac{c_{k}-c_{k}^{u}}{c_{k}^{u}}\right).
\end{equation}
which satisfies $\omega(c_{k},c_{l\neq k}=c^{u}_{l\neq k})= \omega^{u}$
for $c_{k}=c_{k}^{u}$. The important parameter $\kappa_{k}$ denotes the
\emph{impact} that a change of concentration $c_{k}$ has on the
respective transition rate, i.e. it is a measure of \emph{sensitivity}
toward that particular protein. Of course,
$\kappa_{k}=\kappa_{k}(\theta'|\theta)$ in full notation, i.e. the value
does not only change across proteins, but the impact also changes for
different transitions. Putting eqs. (\ref{eq:exp}), (\ref{ansatzexp})
together, we arrive at:
\begin{align}
  \label{raffinierteransatz}
  \Delta\omega_{k} = \omega(c_{k},c_{l\neq k}=c^{u}_{l\neq k})-
  \omega^{u} = \omega^{u} \frac{\kappa_{k}}{c_{k}^{u}} (c_{k}-c_{k}^{u})
\end{align}
In many experimental cases, as e.g. in RNAi screens, the perturbation of
a protein concentration leads to $c_{k}\to 0$, i.e. we are interested in
the dynamics in the \emph{absence} of a given protein. With this
assumption we finally have:
\begin{align}
  \label{final}
  \Delta\omega_{k}(\theta|\theta') = - \kappa_{k}(\theta|\theta')~
  \omega^{u}(\theta|\theta').
\end{align}
This allows us to relate two different dynamical scenarios and their
respective outcome in terms of the vesicle patterns: (i) the
\emph{unperturbed scenario} with experimentally known concentrations and
known transition rates, and (ii) the \emph{perturbed scenario}, where
different proteins may be absent.

As an example, let us investigate how changes in the concentration
$c_{k}$ of the protein $k=\mathrm{CDK8}$ affect the transition rates
$\omega_{\mathrm{CDK8}}(1|0)$ and $\omega_{\mathrm{CDK8}}(0|1)$.
Given the parameters in Figure \ref{fig:patternscdk8}, the reduced
transition rates return
$\Omega_{1}^{\mathrm{CDK8}}-\Omega_{1}^{u}=0.375$, where
$\Omega_{i}^{\mathrm{CDK8}}$ refers to the case where the protein
concentration $c_{k}=c_{\mathrm{CDK8}}=0$, whereas $\Omega_{i}^{u}$ refer
to the unperturbed case. Dividing eqn. (\ref{final}) by
$\omega^{b}(0|1)$, we find
\begin{align}
  \Delta \Omega_{1}^{\mathrm{CDK8}}=
  \Omega_{1}^{\mathrm{CDK8}}-\Omega_{1}^{u}=0.375&=
  -\frac{\omega^{u}(1|0)}{\omega^{b}(0|1)}\cdot
  \kappa_{\mathrm{CDK8}}(1|0)
  \nonumber\\
  &=-\Omega_{1}^{u}\cdot \kappa_{\mathrm{CDK8}}(1|0),
\end{align}
from where it follows that $\kappa_{\mathrm{CDK8}}(1|0)\approx -0.41$,
which describes the sensitivity toward changes in the concentration of
CDK8.

Knowing the difference $ \Delta \Omega_{1}^{\mathrm{CDK8}}$, the
transition rates $\omega_{k}$ are not fully determined because of
$\Omega_{1}^{u}=\omega^{u}(1|0)/\omega^{b}(0|1)$,
$\Omega_{1}^{\mathrm{CDK8}}=\omega_{\mathrm{CDK8}}(1|0)/\omega^{b}(0|1)$. Hence,
we can now discuss two different cases which refer to two hypotheses
about the transport of vesicles towards microtubule minus-ends.

The first hypothesis of our model states that the transition rate
$\omega_{\mathrm{CDK8}}(1|0)$ in cells that are silenced for CDK8 reads:
\begin{equation}
  \label{ansatz}
  \omega_{\mathrm{CDK8}}(1|0)=\omega^{u}(1|0)\cdot
  \exp\bigg(-0.41\cdot\frac{c_{\mathrm{CDK8}}-c_{\mathrm{CDK8}}^{u}}{
    c_{\mathrm{CDK8}}^{u}}\bigg).
\end{equation}
Assuming $c_{\mathrm{CDK8}}\approx 0$, it follows, that
$\omega_{\mathrm{CDK8}}(1|0)$ is increased by a factor of approximately
1.5 with respect to $\omega^{u}(1|0)$. This means that silencing the
protein CDK8 \emph{increases} the transition of vesicles freely diffusing
in the cytosol to vesicles being transported towards the MT-plusends. We
thus hypothesize that CDK8 is either directly or indirectly involved in
the docking of vesicles on microtubule filaments via kinesins.

On the other hand, $\kappa_{\mathrm{CDK8}}(0|1)$ can be obtained by the
following relation:
\begin{align}
  \label{a2}
  \Omega_{1}^{CDK8}-\Omega_{1}^{u}=0.375&=
  \frac{\omega^{b}(1|0)}{\omega^{CDK8}(0|1)}\cdot
  \kappa_{\mathrm{CDK8}}(0|1)\nonumber\\
  &=\Omega_{1}^{CDK8}\cdot \kappa_{\mathrm{CDK8}}(0|1),
\end{align}
from where it follows that $\kappa_{\mathrm{CDK8}}(0|1)\approx 0.29$.
The second hypothesis of our model states that the transition rate
$\omega_{\mathrm{CDK8}}(0|1)$ in cells that are silenced for CDK8 reads:
\begin{equation}
  \label{ansatz}
  \omega_{\mathrm{CDK8}}(0|1)=\omega^{u}(0|1)\cdot
  \exp\bigg(0.29\cdot\frac{c_{\mathrm{CDK8}}-c_{\mathrm{CDK8}}^{u}}{
    c_{\mathrm{CDK8}}^{u}}\bigg).
\end{equation}
Setting $c_{\mathrm{CDK8}}\approx 0$, we find that
$\omega_{\mathrm{CDK8}}(0|1)$ is decreased by a factor of approximately
0.75 with respect to $\omega^{u}(0|1)$. This means that silencing the
protein CDK8 leads to a \emph{decrease} in transition of vesicles
transported towards the MT-plusends to vesicles freely diffusing
vesicles. We thus hypothesize that CDK8 may as well be involved in the
undocking of vesicles bound to MT filaments via kinesins.

In a similar manner, we could estimate the concentration dependence of
other transition rates, using the scaled transition rates
$(\Omega^{CDK8}_{2},\Omega^{CDK8}_{F},\Omega^{CDK8}_{5})$ and
$(\Omega^{u}_{2},\Omega^{u}_{F},\Omega^{u}_{5})$. This enables us to
further develop a number of hypotheses regarding the effect of silencing
the protein CDK8 on the intracellular transport.

\section{Discussion}
\label{conclusions}

\subsection{Motivation of the model}
\label{sec:model}

Genomic and pharmaceutical research nowadays heavily relies on systematic
screens in which the perturbation or silencing of specific proteins
affects the abundance of vesicle patterns observed in a population of
cells. The study of vesicle pattern formation thus is important to
improve our understanding of the function of genes.  To learn how vesicle
patterns are formed, we have set up an agent-based model of intracellular
transport inside a single cell.  Agents represent vesicles which move
either by diffusion in the cytosol or are transported along the
cytoskeletal filaments through molecular motors. Vesicles further
interact with other vesicles by fusion or fission, or they interact with
the cytoskeletal filaments, the cell membrane and the nucleus.  This
interaction is controlled and regulated by specific proteins which are
synthesised inside the cell by transcriptionally active genes. The
activity of these genes thus represents the control parameters of our
system.

Treating vesicles as \emph{Brownian agents} with an internal degree of
freedom allows to formally derive a model that captures all relevant
processes in the formation of vesicle patterns. Five different values of
the internal degree of freedom define the vesicle's different modes of
activity. Transitions between these modes occur as stochastic events
related to the binding of proteins to the vesicle's coat, or to
signalling events. Proteins involved in those processes can control the
vesicle's activity which in turn determines the process of pattern
formation. We therefore assume that the transition rates between
different value of the internal degree of freedom depend on the
concentration of a cell's proteins. To determine the precise transition
rates as functions of protein concentrations would require the full
knowledge about the regulatory structure of gene networks. We simplified
this situation by assuming that the transition rates can be decomposed
into a product of the unperturbed transition rates and an exponential
function depending on the concentration of single proteins. We further
simplified our model by assuming that the cell membrane, the cytoskeletal
filaments and the nucleus are static and that the diffusion coefficient
and velocities along cytoskeletal filaments are constant parameters of
the model. In contrast, the transition rates represent the free
parameters of our model and have been varied.  We have reduced the number
of 10 original transition rates to 4 scaled transition rates. This
introduced an ambiguity in the interpretation of the simulation results
with respect to the original transition rates. We could cope with this
situation by deriving two complementary hypothesis about the influence of
the proteins involved, which could be tested experimentally.

\subsection{Possible comparison with experiments}
\label{sec:exp}

In order to relate our computer simulations to reality, experimental data
are needed to calibrate the simulations. Whenever available, baseline
values for the transition rates obtained from experiments have been
included. This allowed us to generate patterns on real time and spatial
scales. From every simulated pattern, four features can be obtained:
size, relative distance to nucleus, number of vesicles within a fixed
radius around each vesicle and number of vesicles per cell area.
To measure the dissimilarity between the simulated pattern and the
experimentally observed one, it is useful to compute the symmetrized
Kullback-Leibler divergence of the two corresponding vesicle feature
distributions \citep{dissbirbaumer}, to find out for which parameters the
simulated pattern provides a minimum divergence to the experimentally
observed vesicle patterns. An interpretation of these findings in
comparison with hypotheses generated from the computer simulations allows
to draw conclusions about the underlying processes, in particular about
the role of the genes involved.

A comparison of simulated and experimentally measured vesicle patterns
also involves a dimensional problem: our simulations are performed in 2d,
whereas experimental patterns result from vesicle motion in 3d.  In
principle, this would require to correct for the distances as consequence
of the projection from 3d to 2d. Since mammalian cells are relatively
flat with the exception of the nucleus area, we have omitted this
correction. But we considered the fact that vesicles could pass
below/above each other by not assuming mutual exclusion in our 2d
simulations.

\subsection{Future directions}
\label{sec:future}
We emphasize again that our computer simulations lead to testable
hypotheses about the influence of specific genes, such as CDK8, on the
formation of vesicle patterns.  One future application of our model of
intracellular transport concerns the silencing of multiple genes. Let us
denote with $i$ and $l$ two genes which are silenced, then according to
ansatz in eqn.~(\ref{ansatzexp}) we find for the transition rates
\begin{equation}
  \label{ansatzdiscussion}
  \omega_{i;l}(\theta_{j}'|\theta_{j})=\omega(\theta_{j}'|\theta_{j})\cdot
  \exp\bigg(\kappa_{i}(\theta_{j}'|\theta_{j}) 
  \frac{c_{i}-c_{i}^{u}}{c_{i}^{u}} + \kappa_{l}(\theta_{j}'|\theta_{j}) 
  \frac{c_{l}-c_{l}^{u}}{c_{l}^{u}} \bigg).
\end{equation} 
Once we have determined $\kappa_{i}(\theta_{j}'|\theta_{j})$ and
$\kappa_{l}(\theta_{j}'|\theta_{j})$ from single gene silencing
experiments and their related simulations of our model of intracellular
transport, we can predict the resulting pattern and dynamics of the
combined silencing of these two genes.  However,
(eq.~\ref{ansatzdiscussion}) is only valid, if genes $i$ and $l$ are not
in the same (regulatory) pathway. Such a prediction represents an
in-silico experiment and can be tested experimentally.

\subsection*{Acknowledgment}

M.B. could benefit from numerous stimulating discussions with Lucas
Pelkmans.

\end{document}